\documentclass[11pt,a4paper,aip,jmp]{revtex4-1}
\usepackage{amsmath,amssymb,enumerate,epsfig,varioref}

\newtheorem{theorem}{Theorem}[section]

\newtheorem{proposition}{Proposition}[section]

\newtheorem{proof}{Proof}

\def\A{\mathcal{A}}

\def\R{\mathbb{R}}
\def\C{\mathbb{C}}

\def\K{\mathbb{K}}
\def\g{\mathfrak{g}}

\def\p{\partial}
\def\proof{\noindent\textit{Proof. }}
\def\qed{$\blacksquare$}

\def\al{\alpha}
\def\CC{\mathbf{C}}

\def\KK{\mathbf{K}}

\begin{document}

\title{Generalized Heisenberg algebra applied to realizations of the orthogonal, Lorentz and Poincar\'{e} algebras and their dual extensions}

\author{Stjepan Meljanac}
\affiliation{Rudjer Bo\v{s}kovi\'{c} Institute, Theoretical Physics Division, Bijeni\v{c}ka c. 54, HR 10002 Zagreb, Croatia}
\email{meljanac@irb.hr}

\author{Tea Martini\'{c}--Bila\'{c}}
\affiliation{Faculty of Science, University of Split, Rudjera Bo\v{s}kovi\'{c}a 33, 21000 Split, Croatia}
\email{teamar@pmfst.hr}
\email{skresic@pmfst.hr}

\author{Sa\v{s}a Kre\v{s}i\'{c}--Juri\'{c}}
\affiliation{Faculty of Science, University of Split, Rudjera Bo\v{s}kovi\'{c}a 33, 21000 Split, Croatia}

\begin{abstract}
We introduce the generalized Heisenberg algebra $\mathcal{H}_n$ and construct realizations of the orthogonal and Lorentz algebras by
formal power series in a semicompletion of $\mathcal{H}_n$. The obtained realizations are given in terms of the generating function
for the Bernoulli numbers. We also introduce an extension of the orthogonal and Lorentz algebras by quantum angles and study realizations
of the extended algebras in $\mathcal{H}_n$. Furthermore, we show that by extending the generalized Heisenberg algebra $\mathcal{H}_n$ one
can also obtain realizations of the Poincar\'{e} aglebra and its extension by quantum angles.
\end{abstract}


\maketitle

%%%%%%%%%%%%%%%%%%%%%%%%%%%%%%%%%%%%%%%%%%%%%%%%%%%%%%%%%%%%%%%%%%%%%%%%%%%%%%%%%%%%%%%%%%%%%%%%%%%%%%%%%%

\section{Introduction}

Physical evidence suggests that the classical notion of space--time as a continuum is no longer valid at the Planck scale $l_p$ ($l_p = \sqrt{G\hbar /c^3}
\approx 1.62 \times 10^{-35} cm)$. Einstein's theory of gravity coupled with Heisenberg's uncertainty principle implies that space--time
coordinates $\hat x_\mu$ should satisfy uncertainty relations $\Delta \hat x_\mu \Delta \hat x_\nu \geq l_p^2$ (see Refs. \onlinecite{Doplicher-1,Doplicher-2}).
One possible approach towards description of space--time at the Planck scale is in the framework of noncommutave (NC)
geometry based on introducing noncommutativity between space--time coordinates. Algebraic relations satisfied by $\hat x_\mu$ lead to various
models of NC spaces such as the canonical theta--deformed space and Lie algebra type spaces. Realizations of Lie algebras play an important role
in formulation of physical theories on such spaces and in the study of their deformed symmetries. In particular, the $\kappa$--Minkowski space and
$\kappa$-Poincar\'{e} quantum group were studied extensively in Refs. \onlinecite{Lukierski-1, Majid-1, Kowalski-1, Kowalski-2} where the parameter $\kappa$ is usually
interpreted as the Planck mass or quantum gravity scale. The $\kappa$--Poincar\'{e} quantum group represents an example of deformed relativistic
space--time symmetries which lead to deformed dispersion relations. Some related applications can be found in Refs. \onlinecite{Amelino-Camelia, Daszkiewicz, Govindarajan, Govindarajan-2}.
Realizations of NC spaces are based on representing the coordinates $\hat x_\mu$ by a formal power series
in the Heisenberg--Weyl algebra $\mathcal{A}_n$ semicompleted with respect to the degree of a differential operator. Realizations of a large class
of Lie algebra type NC spaces, the associated star--products and their physical applications can be found in Refs. \onlinecite{Durov, Meljanac-1, Meljanac-2,
Meljanac-3, Meljanac-4, Meljanac-5, Meljanac-7, Meljanac-8, Meljanac-9, Meljanac-10, Meljanac-11, Meljanac-12, Harikumar, Meljanac-13}.
Specifically, the Weyl symmetric realization which induces the symmetric ordering on the associated universal enveloping algebra was constructed in
Refs. \onlinecite{Durov, Meljanac-6, Meljanac-14}.

For certain Lie algebras, such as the orthogonal algebra $so(n)$ and Lorentz algebra $so(1,n-1)$, the Weyl symmetric realization is not well adapted due to the
structure of their commutation relations. In the present paper, this is the motivation for introducing the generalized Heisenberg algebra $\mathcal{H}_n$ and
constructing an analogue of the Weyl realization of $so(n)$ and $so(1,n-1)$ by formal power series in a semicompletion of $\mathcal{H}_n$ (for simplicity called
the Weyl realization in $\mathcal{H}_n$). Using a construction of the quantum Poincar\'{e} group related to
$\kappa$--Poincar\'{e} algebra as described in Ref. \onlinecite{Zakrzewski} and duality betweeen them
\cite{Lukierski-2, Lukierski-3, Lukierski-4, Lukierski-5, Kosinski, Lukierski-6},
we obtain an extension of the orthogonal and Lorentz algebras with quantum angles in the limit $\kappa\to \infty$.
Given the Weyl realizations of $so(n)$ and $so(1,n-1)$ in $\mathcal{H}_n$, we also find realizations
of their extensions. Furthermore, we also show that one can find a similar realization of the Poincar\'{e} algebra and its extension by
quantum angles in a suitable extension of the algebra $\mathcal{H}_n$. Using this method, one can also obtain the Weyl realization of the $\kappa$--deformed
Poincar\'{e} algebra \cite{Lukierski-2, Lukierski-3, Lukierski-4, Lukierski-5, Lukierski-6}.

The plan of the paper is as follows. In Section 2 we recall some important facts about the Weyl symmetric realization of a Lie algebra which is needed
in further discussion. We then introduce the generalized Heisenberg algebra $\mathcal{H}_n$ and construct the Weyl realization of $so(n)$ by formal power
series in $\mathcal{H}_n$. The obtained realization is given in terms of the generating function for the Bernoulli numbers. Furthermore, we show that by introducing
the metric tensor of the Minkowski space in the definition of $\mathcal{H}_n$, one can obtain the Weyl realization of the Lorentz algebra $so(1,n-1)$. In Section 3 we extend the
orthogonal and Lorentz algebras by quantum angles and find the realizations of the extended algebras in $\mathcal{H}_n$. We close the section with a brief discussion about
realizations of the Poincar\'{e} algebra and its extension by quantum angles using formal power series in a certain extension of the generalized
Heisenberg algebra $\mathcal{H}_n$.

\section{The Weyl realization of a Lie algebra}

We recall some important facts about the Weyl realization of finite dimensional Lie algebras which is needed in further discussion.
Let $\g$ be a finite dimensional Lie algebra over the field $\K$, ($\K=\R$ or $\K=\C$) with ordered basis $X_1,X_2,\ldots, X_n$
satisfying the commutation relations
\begin{equation}\label{1.01}
[X_\mu,X_\nu]=\sum_{\al=1}^n C_{\mu\nu\al} X_\al.
\end{equation}
The structure constants satisfy $C_{\mu\nu\al} = -C_{\nu\mu\al}$ and the Jacobi identity
\begin{equation}\label{Jacobi}
\sum_{\rho=1}^n \big(C_{\mu\al\rho}\, C_{\rho\beta\nu}+C_{\al\beta\rho}\, C_{\rho\mu\nu} + C_{\beta\mu\rho}\, C_{\rho\al\nu}\big)=0.
\end{equation}
An important example of a Lie algebra type NC space is the $\kappa$--deformed Euclidean space defined by the commutation relations
\begin{equation}\label{1.02}
[X_\mu,X_\nu]=i(a_\mu X_\nu - a_\nu X_\mu), \quad 1\leq \mu,\nu\leq n.
\end{equation}
Here, $a_\mu=v_\mu/\kappa$ where $\kappa\in \R$ is a deformation parameter and $v\in \R^n$ is a unit vector. The algebra \eqref{1.02} was introduced in
Refs. \onlinecite{Lukierski-1} and \onlinecite{Majid-1} and has applications in doubly special relativity
theories \cite{Kowalski-1,Kowalski-2}, quantum gravity \cite{Amelino-Camelia} and quantum field theory \cite{Daszkiewicz, Govindarajan}.
If $\g$ is any Lie algebra defined by Eq. \eqref{1.01}, then by rescaling
the structure constants by a parameter $h\in \R$, $C_{\mu\nu\lambda}\mapsto h C_{\mu\nu\lambda}$, one can think of $\g$ as being a deformation of the
underlying commutative space since $X_\mu X_\nu = X_\nu X_\mu$ as $h\to 0$. Thus, it is of interest to study realizations of $X_\mu$ as deformations
of commutative coordinates $x_\mu$. Such realizations are naturally constructed as embeddings of $\g$ into a semicompletion of the Heisenberg--Weyl algebra $\A_n$
with respect to the degree of a differential operator \cite{Durov}. Recall that $\A_n$ is a unital, associative algebra generated by
$x_\mu$, $\p_\mu$, $1\leq \mu\leq n$, satisfying the commutation relations
\begin{equation}\label{1.04}
[x_\mu,x_\nu]=[\p_\mu,\p_\nu]=0, \quad [\p_\mu,x_\nu]=\delta_{\mu\nu}.
\end{equation}
The realizations considered here are given by
\begin{equation}\label{1.05}
\hat X_\mu = \sum_{\al=1}^n x_\al \varphi_{\al\mu}(\p)
\end{equation}
where $\varphi_{\al\mu}(\p)$ is a formal power series in $\p_1, \p_2, \ldots, \p_n$ depending on the deformation parameter $h$
such that $\lim_{h\to 0} \varphi_{\al\mu}(\p)=\delta_{\al\mu}$. This implies that in the classical limit we have $\lim_{h\to 0}\hat X_\mu = x_\mu$.
The analytic functions $\varphi_{\al\mu}(\p)$ satisfy a system of coupled partial differential equations determined by the commutation relations in $\g$.
Such systems are usually under--determined and have an infinite family of solutions parameterized by arbitrary analytic functions. A number of realizations
of different NC spaces, such as the $\kappa$--deformed space, generalized $\kappa$--deformed space and $su(2)$--type NC space, have been found in Refs.
\onlinecite{Meljanac-1, Meljanac-2, Meljanac-3, Meljanac-4, Meljanac-5}.

To each realization \eqref{1.05} one can associate an ordering on the enveloping algebra $U(\g)$ by using an action of the algebra $\A_n$ on the space
of polynomials $V=\K[x_1,x_2,\ldots, x_n]$ (see Ref. \onlinecite{Meljanac-6}). The action $\rhd \colon \A_n\otimes V\to V$ is defined by
\begin{equation}
x_\mu \rhd f = x_\mu f, \quad \p_\mu\rhd f = \frac{\p f}{\p x_\mu}
\end{equation}
and $(ab)\rhd f = a\rhd (b\rhd f)$ for all $a,b\in \A_n$. Of particular interest is the Weyl symmetric realization associated with the Weyl symmetric
ordering on $U(\g)$. This realization is characterized by the property that
\begin{equation}\label{Weyl_property}
\Big(\sum_{\mu=1}^n k_\mu \hat X_\mu\Big)^m \rhd 1 = \big(\sum_{\mu=1}^n k_\mu x_\mu\big)^m, \quad k_\mu\in \K, \quad m\geq 1,
\end{equation}
or, equivalently, $e^{k\cdot \hat X} \rhd 1 = e^{k\cdot x}$ for all $k\in \K^n$. It has been shown in Refs. \onlinecite{Durov} and \onlinecite{Meljanac-6}
that the Weyl symmetric realization satisfying property \eqref{Weyl_property} can be constructed as follows. Let $\CC=[\CC_{\mu\nu}]$ be the matrix of
differential operators
\begin{equation}\label{1.06A}
\CC_{\mu\nu} = \sum_{\al=1}^n C_{\mu\al\nu} \p_\al
\end{equation}
and let
$\psi(t)$ denote the generating function of the Bernoulli numbers $B_k$,
\begin{equation}\label{1.06}
\psi(t)=\frac{t}{1-e^{-t}}=\sum_{k=0}^\infty \frac{(-1)^k}{k!} B_k t^k.
\end{equation}
Then the symmetric realization of $X_\mu$ is given by
\begin{equation}\label{1.07}
\hat X_\mu = \sum_{\al=1}^n x_\al\, \psi(\CC)_{\mu\al} = \sum_{\al=1}^n x_\al\left(\frac{\CC}{1-e^{-\CC}}\right)_{\mu\al}.
\end{equation}
For odd indices the Bernoulli numbers are $B_1=-\frac{1}{2}$ and $B_{2k+1}=0$ for $k\geq 1$, hence $\psi(t)$ has only even powers of $t$ except for the
lowest order term $\frac{1}{2}t$.

For certain Lie algebras the symmetric realization \eqref{1.07} can be expressed in closed form. For example, for the $\kappa$--deformed space \eqref{1.02}
it is given by \cite{Meljanac-2}
\begin{equation}\label{1.08}
\hat X_\mu = x_\mu \frac{A}{e^A-1} + ia_\mu (x\cdot \p) \frac{e^A-A-1}{(e^A-1)A}
\end{equation}
where $A=i\sum_{k=1}^n a_k \p_k$ and $x\cdot \p = \sum_{k=1}^n x_k \p_k$.

\subsection{Realization of the orthogonal algebra $so(n)$}

In the following we consider realization of the orthogonal algebra $so(n)$ with standard basis $\{M_{\mu\nu}\mid 1\leq \mu< \nu \leq n\}$ satisfying the commutation relations
\begin{equation}\label{1.09}
[M_{\mu\nu},M_{\lambda\rho}]=\delta_{\nu\lambda} M_{\mu\rho} - \delta_{\mu\lambda} M_{\nu\rho} - \delta_{\nu\rho} M_{\mu\lambda} + \delta_{\mu\rho} M_{\nu\lambda}
\end{equation}
where $M_{\mu\nu}=-M_{\nu\mu}$. For future reference let $N=\frac{n(n-1)}{2}$ denote the dimension of $so(n)$.
Unfortunately, the Weyl symmetric realization \eqref{1.07} cannot be applied directly to rotation generators $M_{\mu\nu}$ since this requires an
ordering on the set $\{M_{\mu\nu}\mid 1\leq \mu<\nu \leq n\}$ in order to establish a correspondence between $M_{\mu\nu}$ and
the generators of the Heisenberg--Weyl algebra $\A_N$. We show that it is more natural to consider a realization of $M_{\mu\nu}$ using formal power series in the
generalized Heisenberg algebra $\mathcal{H}_n$ defined as follows. The algebra $\mathcal{H}_n$ is a unital, associative algebra generated by $x_{\mu\nu}$, $\p_{\mu\nu}$,
$1\leq \mu,\nu\leq n$, satisfying $x_{\mu\nu}=-x_{\nu\mu}$, $\p_{\mu\nu}=-\p_{\nu\mu}$ and commutation relations
\begin{equation}\label{Hn}
[x_{\mu\nu},x_{\al\beta}]=0, \quad [\p_{\mu\nu},\p_{\al\beta}]=0 \quad \text{and}\quad [\p_{\mu\nu},x_{\al\beta}]=\delta_{\mu\al} \delta_{\nu\beta}-
\delta_{\mu\beta} \delta_{\nu\al}.
\end{equation}
The idea is to apply the symmetric realization \eqref{1.07} to an isomorphic image of $so(n)$ and then use the inverse map in order to obtain
the realization of $M_{\mu\nu}$ in the algebra $\mathcal{H}_n$. At this point it is useful to introduce the following convention.
The greek indices $\al,\beta,\gamma, \ldots$ run through the set $\{1,2,\ldots, n\}$ and the latin indices $a,b,c\ldots$ run
through the set $\{1,2,\ldots, N\}$. The commutation relations for $so(n)$ can be written as
\begin{equation}
[M_{\mu\nu},M_{\lambda\rho}]=\sum_{\al,\beta=1}^n C_{(\mu\nu)(\lambda\rho)(\al\beta)}\, M_{\al\beta}
\end{equation}
where the structure constants are given by
\begin{align}
C_{(\mu\nu)(\lambda \rho) (\al \beta)} &= \frac{1}{2}(\delta_{\mu\al} \delta_{\rho\beta}-\delta_{\mu\beta}\delta_{\rho\al}) \delta_{\nu\lambda} -
\frac{1}{2}( \delta_{\nu\al} \delta_{\rho\beta}-\delta_{\nu\beta}\delta_{\rho\al}) \delta_{\mu\lambda} \notag \\
&+\frac{1}{2}(\delta_{\lambda\al} \delta_{\mu\beta}-\delta_{\lambda\beta}\delta_{\mu\al}) \delta_{\nu\rho}
-\frac{1}{2}(\delta_{\lambda\al} \delta_{\nu\beta}-\delta_{\lambda\beta}\delta_{\nu\al}) \delta_{\mu\rho}.  \label{CC}
\end{align}
Let us define
\begin{equation}\label{1.14}
M_a = \frac{1}{2} \sum_{\mu,\nu=1}^n \Gamma_a^{\mu\nu}\, M_{\mu\nu}, \quad a=1,2,\ldots, N,
\end{equation}
where the coefficients $\Gamma_a^{\mu\nu}$ satisfy $\Gamma_a^{\mu\nu}=-\Gamma_a^{\nu\mu}$. The inverse transformation is given by
\begin{equation}\label{1.15}
M_{\mu\nu} = \sum_{a=1}^N \Gamma_{\mu\nu}^a\, M_a
\end{equation}
where $\Gamma_{\mu\nu}^a = -\Gamma_{\nu\mu}^a$. In view of Eqs. \eqref{1.14} and \eqref{1.15} the transformation coefficients are coupled
by nonlinear relations
\begin{equation}\label{1.17}
\frac{1}{2} \sum_{\mu,\nu=1}^n \Gamma_{\mu\nu}^a \, \Gamma_b^{\mu\nu} = \delta_{ab} \quad \text{and} \quad \sum_{a=1}^N
\Gamma_{\mu\nu}^a\, \Gamma_a^{\al\beta} =
\delta_{\mu\al} \delta_{\nu\beta} - \delta_{\mu\beta} \delta_{\nu\al}.
\end{equation}
We note that there is a certain amount of symmetry involved in the coefficients $\Gamma_a^{\mu\nu}$ and $\Gamma_{\mu\nu}^a$ which is important for
later discussion. If we multiply the first
relation in Eq. \eqref{1.17} by $\Gamma_b^{\al\beta}$ and then use the second relation, we find $\Gamma_{\al\beta}^a = \Gamma_a^{\al\beta}$. Thus, we
also have
\begin{align}
\frac{1}{2} \sum_{\mu,\nu=1}^n \Gamma_{\mu \nu}^a \Gamma_{\mu \nu}^b &= \frac{1}{2}\sum_{\mu,\nu=1}^n \Gamma_a^{\mu \nu} \Gamma_b^{\mu\nu} = \delta_{ab}  \label{1.19B}
\intertext{and similarly}
\sum_{a=1}^N \Gamma_{\mu\nu}^a \Gamma_{\al\beta}^a &= \sum_{a=1}^N \Gamma_a^{\mu\nu} \Gamma_a^{\al\beta} = \delta_{\mu\al} \delta_{\nu\beta} -
\delta_{\mu\beta} \delta_{\nu\al}.  \label{1.20B}
\end{align}
We demand that the linear transformation \eqref{1.14} defines a Lie algebra isomorphism. Hence, we require that $M_a$ generate a Lie algebra defined by commutation
relations
\begin{equation}
[M_a,M_b]=\sum_{c=1}^N C_{abc}\, M_c
\end{equation}
with structure constants
\begin{align}
C_{abc} &= \frac{1}{4} \sum_{\mu,\nu=1}^n \sum_{\lambda,\rho=1}^n \sum_{\al,\beta=1}^n \Gamma_a^{\mu\nu}\, \Gamma_b^{\lambda\rho}\,\Gamma_{\al\beta}^c\,
C_{(\mu \nu) (\lambda \rho) (\al\beta)}  \label{Cabc} \\
&=\frac{1}{2}\sum_{\al,\beta,\lambda=1}^n \big(\Gamma_a^{\al\lambda} \Gamma_b^{\lambda\beta}-\Gamma_b^{\al\lambda} \Gamma_a^{\lambda\beta}\big) \Gamma^c_{\al\beta}. \label{1.19}
\end{align}
It is easy to see that for the orthogonal algebra $so(3)$ we have $\Gamma_a^{\mu\nu}=\varepsilon_{a\mu\nu}$ and $C_{abc}=\varepsilon_{abc}$ where $\varepsilon_{abc}$ is the Levi--Civita
symbol. Let us define a transformation $(x_{\mu\nu},\p_{\mu\nu})\to (x_a,\p_a)$ by
\begin{equation}\label{xapa}
x_a = \frac{1}{2}\sum_{\mu,\nu=1}^n \Gamma_a^{\mu\nu} x_{\mu\nu}, \quad \p_a = \frac{1}{2}\sum_{\mu,\nu=1}^n \Gamma_a^{\mu\nu} \p_{\mu\nu}, \quad 1\leq a \leq N.
\end{equation}
Then the inverse transformation is given by
\begin{equation}\label{invxd}
x_{\mu\nu} = \sum_{a=1}^N \Gamma_{\mu\nu}^a x_a, \quad \p_{\mu\nu} = \sum_{a=1}^N \Gamma_{\mu\nu}^a \p_a, \quad 1\leq \mu,\nu\leq n.
\end{equation}
The identities satisfied by the coefficients $\Gamma_a^{\mu\nu}$ and $\Gamma_{\mu\nu}^a$ imply that $x_a$ and $\p_a$ close the Heisenberg--Weyl algebra $\A_N$,
\begin{equation}\label{1.26B}
[x_a,x_b]=0, \quad [\p_a,\p_b]=0, \quad [\p_a,x_b]=\delta_{ab}.
\end{equation}
Indeed, the first two relations in \eqref{1.26B} follow trivially while the third relation is a direct consequence of Eq. \eqref{1.19B}. Therefore, we
can use Eq. \eqref{1.07} to write the symmetric realization of $M_a$,
\begin{equation}\label{1.24}
\hat M_a = \sum_{b=1}^N x_b\, \psi(\CC)_{ab}
\end{equation}
where $\CC_{ab} = \sum_{c=1}^N C_{acb} \p_c$ and $C_{abc}$ are the structure constants given by Eq. \eqref{1.19}. Since $C_{acb}$
depend on the coefficients $\Gamma^{\mu\nu}_a$ which in turn define
the isomorphism of $so(n)$, so does the realization \eqref{1.24}. Our goal is to transform the realization \eqref{1.24} into a realization of the
standard rotation generators $M_{\mu\nu}$ using the generalized Heisenberg algebra $\mathcal{H}_n$. This realization will be independent of
the transformation coefficients $\Gamma_a^{\mu\nu}$.

Let us express the matrix elements $\CC_{ab}$ in terms of the generators $\p_{\mu\nu}$. It follows from Eqs. \eqref{Cabc} and \eqref{xapa} that
\begin{equation}
\CC_{ab}=\frac{1}{4} \sum_{\theta,\sigma=1}^n \sum_{\mu,\nu=1}^n \sum_{\lambda,\rho=1}^n \sum_{\al,\beta=1}^n \Gamma_a^{\mu\nu}\, \Gamma_{\al\beta}^b\, C_{(\mu\nu)(\lambda\rho)(\al\beta)}\,
\frac{1}{2}\Big(\sum_{c=1}^N \Gamma_c^{\lambda\rho}\, \Gamma_c^{\theta\sigma}\Big) \p_{\theta\sigma}.
\end{equation}
Since $\Gamma_a^{\mu\nu}$ satisfy condition \eqref{1.19B}, the above expression can be simplified as
\begin{equation}
\CC_{ab} = \frac{1}{4} \sum_{\theta,\sigma=1}^n \sum_{\mu,\nu=1}^n \sum_{\al,\beta=1}^n \Gamma_a^{\mu\nu}\, \Gamma_{\al\beta}^b\, C_{(\mu\nu)(\theta\sigma)(\al\beta)}\, \p_{\theta\sigma}.
\end{equation}
If we denote
\begin{equation}\label{KK}
\KK_{(\mu\nu)(\al\beta)} = \frac{1}{2} \sum_{\theta,\sigma=1}^n C_{(\mu\nu)(\theta\sigma)(\al\beta)}\, \p_{\theta\sigma}
\end{equation}
in analogy with differential operators $\CC_{\mu\nu}$ defined by Eq. \eqref{1.06A}, then
\begin{equation}\label{Cab}
\CC_{ab} = \frac{1}{2} \sum_{\mu,\nu=1}^n \sum_{\al,\beta=1}^n \Gamma_a^{\mu\nu}\, \Gamma_{\al\beta}^b\, \KK_{(\mu\nu)(\al\beta)}.
\end{equation}
Since the structure constants $C_{(\mu\nu)(\lambda\rho)(\al\beta)}$ are given explicitly by Eq. \eqref{CC},
the elements $\KK_{(\mu\nu)(\al\beta)}\in \mathcal{H}_n$ are given by
\begin{equation}\label{1.30A}
\KK_{(\mu\nu)(\al\beta)} = \frac{1}{2} (\delta_{\mu\al}\, \p_{\nu\beta} - \delta_{\mu\beta}\, \p_{\nu\al}+\delta_{\nu\beta}\,
\p_{\mu\al} - \delta_{\nu\al}\, \p_{\mu\beta}).
\end{equation}
We can organize them into an $n^2\times n^2$ matrix $\KK=[\KK_{(\mu\nu)(\al\beta)}]$ for a fixed ordering
of the pairs $(\mu,\nu)$ and $(\al,\beta)$, respectively. The powers of $\KK$ are given recursively by
\begin{equation}\label{1.32A}
(\KK^m)_{(\mu\nu)(\al\beta)} = \sum_{\theta,\sigma=1}^n \KK_{(\mu\nu) (\theta\sigma)}\, (\KK^{m-1})_{(\theta\sigma)(\al\beta)} =
\sum_{\theta,\sigma=1}^n (\KK^{m-1})_{(\mu\nu) (\theta\sigma)}\, \KK_{(\theta\sigma)(\al\beta)} \quad m\geq 1,
\end{equation}
where by definition
\begin{equation}\label{1.33}
(\KK^0)_{(\mu\nu)(\al\beta)} = \frac{1}{2}(\delta_{\mu\al}\, \delta_{\nu\beta} - \delta_{\mu\beta}\, \delta_{\nu\al}).
\end{equation}
We note that the powers of $\KK$ are skew--symmetric with respect to transposition of indices, i.e. $(\KK^m)_{(\mu\nu)(\al\beta)}=-(\KK^m)_{(\nu\mu)(\al\beta)}$
and $(\KK^m)_{(\mu\nu)(\al\beta)}=-(\KK^m)_{(\mu\nu)(\beta\al)}$. This follows immediately from the definition of $\KK_{(\mu\nu)(\al\beta)}$ and
Eq. \eqref{1.32A}. The matrix $\KK$ plays a completely analogous role to that of the matrix $\CC$ in the realization of the rotation generators $M_{\mu\nu}$.
In constructing this realization we need the following transformation between the matrices $\psi(\CC)$ and $\psi(\KK)$.

\begin{proposition}\label{prop-1}
\begin{equation}
\psi(\CC)_{ab} = \frac{1}{2} \sum_{\mu,\nu=1}^n \sum_{\al,\beta=1}^n \Gamma_a^{\mu\nu}\, \Gamma^b_{\al\beta}\, \psi(\KK)_{(\mu\nu)(\al\beta)}
\end{equation}
where $\psi(t)$ is the generating function for the Bernoulli numbers \eqref{1.06}.
\end{proposition}

\proof First we prove by induction that
\begin{equation}\label{1.32}
(\CC^m)_{ab} = \frac{1}{2} \sum_{\mu,\nu=1}^n \sum_{\al,\beta=1}^n \Gamma_a^{\mu\nu}\, \Gamma^b_{\al\beta}\, (\KK^m)_{(\mu\nu) (\al\beta)}, \quad m\geq 0.
\end{equation}
Since $(\KK^0)_{(\mu\nu)(\al\beta)}=\frac{1}{2}(\delta_{\mu\al}\, \delta_{\nu\beta}-\delta_{\mu\beta}\, \delta_{\al\beta})$, a short computation using the
first identity in Eq. \eqref{1.17} shows that the claim holds for $m=0$. Now suppose that Eq. \eqref{1.32} is true for some $m>0$. Then using
Eq. \eqref{Cab} we have
\begin{align}
(\CC^{m+1})_{ab} &= \sum_{c=1}^N \CC_{ac} (\CC^m)_{cb} \notag \\
&= \sum_{c=1}^N \Big(\frac{1}{2} \sum_{\mu,\nu=1}^n \sum_{\rho,\sigma=1}^n \Gamma_a^{\mu\nu} \, \Gamma^c_{\rho\sigma}\, \KK_{(\mu\nu)(\rho\sigma)}\Big)
\Big(\frac{1}{2} \sum_{\theta,\tau=1}^n \sum_{\al,\beta=1}^n \Gamma_c^{\theta\tau}\, \Gamma^b_{\al\beta}\, (\KK^m)_{(\theta\tau)(\al\beta)}\Big). \label{1.34}
\end{align}
We note that the second identity in Eq. \eqref{1.17} yields
\begin{align}
&\sum_{\rho,\sigma=1}^n \sum_{\theta,\tau=1}^n\Big(\sum_{c=1}^N \Gamma_{\rho\sigma}^c\, \Gamma_c^{\theta\tau}\Big)\, \KK_{(\mu\nu)(\rho\sigma)} (\KK^m)_{(\theta\tau)(\al\beta)} \notag \\
&=\sum_{\rho,\sigma=1}^n \sum_{\theta,\tau=1}^n \big(\delta_{\theta\rho}\, \delta_{\tau\sigma} - \delta_{\theta\sigma}\, \delta_{\tau\rho}\big) \KK_{(\mu\nu)(\rho\sigma)}
(\KK^m)_{(\theta\tau)(\al\beta)} \notag \\
&= \sum_{\rho,\sigma=1}^n \Big[\KK_{(\mu\nu)(\rho\sigma)} (\KK^m)_{(\rho\sigma)(\al\beta)} - \KK_{(\mu\nu)(\rho\sigma)} (\KK^m)_{(\sigma\rho)(\al\beta)}\Big] \notag \\
&= 2 \sum_{\rho,\sigma=1}^n \KK_{(\mu\nu) (\rho\sigma)} (\KK^m)_{(\rho\sigma)(\al\beta)} = 2 (\KK^{m+1})_{(\mu\nu)(\al\beta)}  \label{1.35}
\end{align}
where we have taken into account that $(\KK^m)_{(\mu\nu)(\al\beta)} = - (\KK^m)_{(\nu\mu)(\al\beta)}$ for $m\geq 1$. Now, substituting Eq. \eqref{1.35} into Eq. \eqref{1.34}
we find that
\begin{equation}
(\CC^{m+1})_{ab} = \frac{1}{2} \sum_{\mu,\nu=1}^n \sum_{\al,\beta=1}^n \Gamma_a^{\mu\nu}\, \Gamma^b_{\al\beta}\, (\KK^{m+1})_{(\mu\nu)(\al\beta)}
\end{equation}
which proves the claim. Inserting Eq. \eqref{1.32} into the generating function for the Bernoulli numbers \eqref{1.06} we obtain
\begin{equation}
\psi(\CC)_{ab} = \sum_{k=0}^\infty \frac{(-1)^k}{k!} B_k\, (\CC^k)_{ab} = \frac{1}{2} \sum_{\mu,\nu=1}^n \sum_{\al,\beta=1}^n \Gamma_a^{\mu\nu}\, \Gamma^b_{\al\beta}\,
\psi(\KK)_{(\mu\nu)(\al\beta)}.
\end{equation}
\qed

Now we can prove the key result about the Weyl realization of the rotation algebra $so(n)$.

\begin{theorem}\label{tm-1}
Realization of the rotation generators $M_{\mu\nu}$ by formal power series in the generalized Heisenberg algebra $\mathcal{H}_n$ is given by
\begin{equation}\label{1.40C}
\hat M_{\mu\nu} = \sum_{\al,\beta=1}^n x_{\al\beta}\, \psi(\KK)_{(\mu\nu) (\al\beta)}.
\end{equation}
\end{theorem}

\proof Recall that the Weyl symmetric realization of the generators $M_a$ is given by Eq. \eqref{1.24}. Hence, it follows from transformation \eqref{1.15} that
a realization of $M_{\mu\nu}$ in the Heisenberg--Weyl algebra $\A_N$ is given by
\begin{equation}
\hat M_{\mu\nu} = \sum_{a=1}^N \sum_{b=1}^N \Gamma_{\mu\nu}^a\, x_b\, \psi(\CC)_{ab}.
\end{equation}
Since $x_b$ is defined by Eq. \eqref{xapa} and $\psi(\CC)_{ab}$ is given by Proposition \ref{prop-1}, the realization of $M_{\mu\nu}$ in the
generalized Heisenberg algebra $\mathcal{H}_n$ can be written as
\begin{equation}
\hat M_{\mu\nu} = \frac{1}{4} \sum_{\al,\beta=1}^n x_{\al \beta} \sum_{\rho,\sigma=1}^n \sum_{\tau,\theta=1}^n \Big(\sum_{a=1}^N \Gamma_{\mu\nu}^a \Gamma_a^{\rho\sigma}\Big)
\Big(\sum_{b=1}^N \Gamma_{\tau\theta}^b \Gamma_b^{\al\beta}\Big)\psi(\KK)_{(\rho\sigma)(\tau\theta)}.
\end{equation}
One can simplify the above expression by using the second identity in Eq. \eqref{1.17} which leads to
\begin{equation}
\hat M_{\mu\nu} = \frac{1}{4} \sum_{\al,\beta=1}^n x_{\al\beta}\, \Big(\psi(\KK)_{(\mu\nu)(\al\beta)}-\psi(\KK)_{(\nu\mu)(\al\beta)}-\psi(\KK)_{(\mu\nu)(\beta\al)}+
\psi(\KK)_{(\nu\mu)(\beta\al)}\Big).
\end{equation}
In view of the skew--symmetric property $\psi(\KK)_{(\mu\nu)(\al\beta)}=-\psi(\KK)_{(\nu\mu)(\al\beta)}$ and $x_{\al\beta}=-x_{\beta\al}$, the last expression
becomes
\begin{equation}\label{1.42}
\hat M_{\mu\nu} = \sum_{\al,\beta=1}^n x_{\al\beta}\, \psi(\KK)_{(\mu\nu)(\al\beta)}.
\end{equation}
\qed\\
We note that realization \eqref{1.40C} is formally similar to realization \eqref{1.24}, but it uses a different algebra to describe the rotation generators
of $so(n)$. We close this section by stating that the powers of $\KK$ appearing in the formal power series of the generating function
\begin{equation}
\psi(\KK)_{(\mu\nu)(\al\beta)} = \sum_{m=0}^\infty \frac{(-1)^m}{m!} B_m\, (\KK^m)_{(\mu\nu)(\al\beta)}
\end{equation}
are polynomials in $\p_{\mu\nu}$ explicitly given by
\begin{equation}\label{1.45}
(\KK^m)_{(\mu\nu)(\al\beta)} = \frac{1}{2}\sum_{k=0}^m \binom{m}{k} \big(\p_{\mu\al}^k \p_{\nu\beta}^{m-k}-
\p_{\mu\beta}^{m-k} \p_{\nu\al}^k\big), \quad m\geq 0
\end{equation}
(for proof see Proposition \ref{A-1} in the  Appendix).

\subsection{Realization of the Lorentz algebra $so(1,n-1)$}

In physical applications one is frequently interested in the Lorentz algebra $so(1,n-1)$ defined by
\begin{equation}\label{1.47A}
[M_{\mu\nu},M_{\lambda\rho}]=\eta_{\nu\lambda} M_{\mu\rho} - \eta_{\mu\lambda} M_{\nu\rho} - \eta_{\nu\rho} M_{\mu\lambda} + \eta_{\mu\rho} M_{\nu\lambda}
\end{equation}
where $\eta=diag(-1,1,\ldots, 1)$ is the Minkowski metric. To keep the notation consistent with previous sections, we let the greek indices run through
the values $1,2,\ldots, n$ instead of the more common values $0,1,\ldots, n-1$. The same applies to the Poincar\'{e} algebra in Section 3.
Our discussion presented in the previous subsection can be easily adapted to give the corresponding realization
of the algebra \eqref{1.47A}. In this case we must make the following modifications. The generalized Heisenberg algebra, here denoted $\tilde{\mathcal{H}}_n$, is
defined by commutation relations
\begin{equation}\label{1.48A}
[x_{\mu\nu},x_{\al\beta}]=0, \quad [\p_{\mu\nu},\p_{\al\beta}]=0 \quad \text{and}\quad [\p_{\mu\nu},x_{\al\beta}]=\eta_{\mu\al}\, \eta_{\nu\beta}-
\eta_{\mu\beta}\, \eta_{\nu\al}.
\end{equation}
The elements of the matrix $\KK=[\KK_{(\mu\nu)(\al\beta)}]$ are defined by
\begin{equation}
\KK_{(\mu\nu) (\al\beta)} = \frac{1}{2} (\eta_{\mu\al}\, \p_{\nu\beta} - \eta_{\mu\beta}\, \p_{\nu\al}+\eta_{\nu\beta}\, \p_{\mu\al}-\eta_{\nu\al}\, \p_{\mu\beta})
\end{equation}
and the powers of $\KK$ are given recursively by
\begin{align}
(\KK^m)_{(\mu\nu)(\al\beta)} &= \sum_{\lambda,\rho=1}^n \sum_{\lambda^\prime, \rho^\prime=1}^n \KK_{(\mu\nu)(\lambda\rho)}\, \eta_{\lambda \lambda^\prime}\, \eta_{\rho\rho^\prime}\,
(\KK^{m-1})_{(\lambda^\prime \rho^\prime)(\al\beta)} \\
&=\sum_{\lambda,\rho=1}^n \sum_{\lambda^\prime, \rho^\prime=1}^n (\KK^{m-1})_{(\mu\nu)(\lambda\rho)}\, \eta_{\lambda \lambda^\prime}\, \eta_{\rho\rho^\prime}\,
\KK_{(\lambda^\prime \rho^\prime)(\al\beta)}
\end{align}
where
\begin{equation}
(\KK^0)_{(\mu\nu)(\al\beta)} = \frac{1}{2}(\eta_{\mu\al}\, \eta_{\nu\beta}-\eta_{\mu\beta}\, \eta_{\nu\al}).
\end{equation}
Then the the Lorentz algebra \eqref{1.47A} admits the realization
\begin{equation}\label{2.53-B}
\hat M_{\mu\nu} = \sum_{\al,\beta=1}^n \sum_{\al^\prime, \beta^\prime=1}^n x_{\al\beta}\, \eta_{\al\al^\prime}\, \eta_{\beta\beta^\prime}\, \psi(\KK)_{(\mu\nu)(\al^\prime \beta^\prime)}
\end{equation}
where $\psi(t)$ is the generating function for the Bernoulli numbers \eqref{1.06}.

\section{Realizations of the orthogonal and Lorentz algebras extended by quantum angles}

Recently there has been a growth of interest in studying the bialgebroid and Hopf algebroid structure of deformed quantum phase spaces with
noncommutative coordinates (see Refs. \onlinecite{Lukierski-2, Lukierski-3, Lukierski-4, Lukierski-5, Kosinski, Lukierski-6}).
In Ref. \onlinecite{Lukierski-2} the authors study the generalized quantum phase space
$\mathcal{H}^{(10,10)}=(\hat \xi_\mu, \hat \Lambda_{\mu\nu}, P_\mu,M_{\mu\nu})$ where $\mathbb{H}=(P_\mu,M_{\mu\nu})$ is the classical Poincar\'{e}--Hopf
algebra and $\hat G=(\hat \xi_\mu, \hat \Lambda_{\mu\nu})$ is the quantum Poincar\'{e} group dual to $\mathbb{H}$. Here $\hat \xi_\mu$ denote quantum translations and
$\hat \Lambda_{\mu\nu}$ represent
quantum rotations dual to the generators $M_{\mu\nu}$. The quantum Poincar\'{e} group $\hat G$ is roughly constructed as follows
(for more details see Ref. \onlinecite{Lukierski-2}).
One associates to $\mathbb{H}$ the primitive Hopf algebra structure and uses a family of Abelian twists $\mathcal{F}_u\in \mathbb{H}\otimes \mathbb{H}$
to deform the coalgebra sector of $\mathbb{H}$. The universal $\mathcal{R}$--matrix $\mathcal{R}=\mathcal{F}^T_u \mathcal{F}^{-1}_u$ and the
Faddeev--Reshetikhin--Takhtajan procedure \cite{Faddeev} is used to construct the quantum Poincar\'{e} group $\hat G$. Duality between $\mathbb{H}$ and
$\hat G$ is defined by establishing a pairing between $\mathbb{H}$ and $\hat G$ which is then used to construct an action of $\mathbb{H}$ on $\hat G$
endowing $\hat G$ with the structure of an $\mathbb{H}$--module. Using the action of $\mathbb{H}$ on $\hat G$, the quantum phase space $\mathcal{H}^{(10,10)}$
is constructed as the Heisenberg double algebra $\mathbb{H} \rtimes \hat G$. If $\hat \Lambda$ denotes the matrix with elements $\hat \Lambda_{\mu\nu}$, then
$\hat \Lambda$ satisfies the group property $\hat \Lambda^T \hat \Lambda = \hat \Lambda \hat \Lambda^T=I$, thus justifying the term ``quantum angles`` for
$\hat \Lambda_{\mu\nu}$. The commutation relations describing the structure of $\mathbb{H}\rtimes \hat G$ are fairly complicated and clearly depend
on deformation parameters of the twist element $\mathcal{F}$.

In this section we are interested in extension of the Lorentz algebra \eqref{1.47A} by quantum angles $\Lambda_{\mu\nu}$ (written without hat) in the limit $\kappa\to \infty$ where $\kappa$ is
a deformation parameter of the twist element $\mathcal{F}$ (cf. Eqs. (43) and (51) in Ref. \onlinecite{Lukierski-2}). In this case $M_{\mu\nu}$ and $\Lambda_{\mu\nu}$ define a pair of undeformed
dual variables in the Lorentz sector which close the Lie algebra defined by relations \eqref{1.47A} and
\begin{align}
[M_{\mu\nu},\Lambda_{\rho\sigma}] &= \eta_{\rho\nu}\, \Lambda_{\mu\sigma} - \eta_{\rho\mu} \Lambda_{\nu\sigma},  \label{3.54A} \\
[\Lambda_{\mu\nu}, \Lambda_{\rho\sigma}] &=0.   \label{3.55A}
\end{align}
We shall use the techniques developed in Sec. 2 to find a realization of the above algebra as well as the extended orthogonal algebra.

First, let us consider the extended rotation algebra $so(n)$ defined by commutation relations \eqref{1.09} and
\begin{align}
[M_{\mu\nu},\Lambda_{\rho\sigma}] &= \delta_{\nu\rho} \Lambda_{\mu\sigma}-\delta_{\mu\rho}\Lambda_{\nu\sigma}, \label{3.76} \\
[\Lambda_{\mu\nu},\Lambda_{\rho\sigma}] &= 0.  \label{3.77}
\end{align}
Given the symmetric realization \eqref{1.40C} of the rotation generators $M_{\mu\nu}$ we seek a realization of the quantum angles
$\Lambda_{\mu\nu}$ by formal power series in the generalized Heisenberg algebra $\mathcal{H}_n$. We assume that the realization of $\Lambda_{\mu\nu}$
depends only on the generators $\p_{\al\beta}\in \mathcal{H}_n$.

\begin{theorem}
The realization of $\Lambda_{\mu\nu}$ is given by the exponential function
\begin{equation}
\hat \Lambda_{\mu\nu}=(e^\p)_{\mu\nu}
\end{equation}
where $\p=[\p_{\al\beta}]$ is the $n\times n$ matrix of generators $\p_{\al\beta}\in \mathcal{H}_n$.
\end{theorem}

\proof We seek a realization of $\Lambda_{\mu\nu}$ in the form
\begin{equation}
\hat \Lambda_{\mu\nu} = \sum_{m=0}^\infty a_m (\p^m)_{\mu\nu}.
\end{equation}
Substituting realization
\eqref{1.40C} into the commutation relation \eqref{3.76} we find that
\begin{equation}\label{3.78}
\sum_{\al,\beta=1}^n [x_{\al\beta},\hat \Lambda_{\rho\sigma}]\, \psi(\KK)_{(\mu\nu)(\al\beta)} = \delta_{\rho\nu}\, \hat \Lambda_{\mu\sigma} -
\delta_{\rho\mu}\, \hat \Lambda_{\nu\sigma}.
\end{equation}
The inverse of the matrix $\psi(\KK)$ is given by the power series
\begin{equation}
\psi(\KK)^{-1}_{(\mu\nu)(\al\beta)} = \left(\frac{1-e^{-\KK}}{\KK}\right)_{(\mu\nu)(\al\beta)} = \sum_{m=1}^\infty \frac{1}{m!} (-\KK)^{m-1}_{(\mu\nu)(\al\beta)}
\end{equation}
and satisfies the identity
\begin{equation}\label{3.80}
\sum_{\al,\beta=1}^n \psi(\KK)_{(\mu\nu)(\al\beta)}\, \psi(\KK)^{-1}_{(\al\beta)(\lambda\rho)} = \frac{1}{2}(\delta_{\mu\lambda}\, \delta_{\nu\rho}-
\delta_{\mu\rho}\, \delta_{\nu\lambda}).
\end{equation}
Hence, using Eq. \eqref{3.80} we can write Eq. \eqref{3.78} as
\begin{equation}\label{3.83}
[x_{\al\beta},\hat \Lambda_{\rho\sigma}] = 2 \sum_{\mu=1}^n \psi(\KK)^{-1}_{(\al\beta)(\mu\rho)}\, \hat \Lambda_{\mu\sigma}
\end{equation}
where we have taken into account the antisymmetric property $\psi(\KK)^{-1}_{(\mu\nu)(\al\beta)} = -\psi(\KK)^{-1}_{(\mu\nu)(\beta\al)}$. Substituting the power series
for $\hat \Lambda_{\rho\sigma}$ and $\psi(\KK)^{-1}_{(\al\beta)(\mu\rho)}$ into Eq. \eqref{3.83} we find that the coefficients $a_m$ satisfy
\begin{equation}\label{3.84}
\sum_{m=1}^\infty a_m [x_{\al\beta},(\p^m)_{\rho\sigma}] = 2\sum_{m=0}^\infty \sum_{k=1}^\infty a_m\, \frac{1}{k!} \sum_{\mu=1}^n (-\KK)^{k-1}_{(\al\beta)(\mu\rho)}\,
(\p^m)_{\mu\sigma}.
\end{equation}
A straightforward but lengthy computation (see Proposition \ref{A-3} in the Appendix) shows that the commutation $[x_{\al\beta},(\p^m)_{\rho\sigma}]$
can be written in terms of the powers of $\KK$ as
\begin{equation}
[x_{\al\beta},(\p^m)_{\rho\sigma}] = 2 \sum_{k=1}^m \binom{m}{k} \sum_{\mu=1}^n (-\KK)^{k-1}_{(\al\beta)(\mu\rho)}\, (\p^{m-k})_{\mu\sigma}.
\end{equation}
Combining the above result with Eq. \eqref{3.84} we obtain
\begin{equation}\label{3.86}
\sum_{m=1}^\infty \sum_{k=1}^m a_m \binom{m}{k} \sum_{\mu=1}^n (-\KK)^{k-1}_{(\al\beta)(\mu\rho)}\, (\p^{m-k})_{\mu\sigma} =
\sum_{m=0}^\infty \sum_{k=1}^\infty a_m \frac{1}{k!} \sum_{\mu=1}^n (-\KK)^{k-1}_{(\al\beta)(\mu\rho)}\, (\p^m)_{\mu\sigma}.
\end{equation}
Rearranging the terms on the left--hand side of Eq. \eqref{3.86} it follows that
\begin{equation}
\sum_{m=0}^\infty \sum_{k=1}^\infty a_{m+k} \binom{m+k}{k} \sum_{\mu=1}^n (-\KK)^{k-1}_{(\al\beta)(\mu\rho)} \, (\p^m)_{\mu\sigma} =
\sum_{m=0}^\infty \sum_{k=1}^\infty a_m \frac{1}{k!} \sum_{\mu=1}^n (-\KK)^{k-1}_{(\al\beta)(\mu\rho)}\, (\p^m)_{\mu\sigma}
\end{equation}
which implies that the coefficients $a_m$ satisfy the recurrence relation
\begin{equation}
\frac{a_{m+k}}{a_m} = \frac{m!}{(m+k)!}.
\end{equation}
For $m=0$ we obtain $a_k = a_0/k!$ and after normalizing the realization by taking $a_0=1$ we find $\hat \Lambda_{\mu\nu}=(e^\p)_{\mu\nu}$. \qed\\

\noindent A similar computation can be carried out to find the realization of the extended Lorenz algebra \eqref{3.54A}--\eqref{3.55A} by formal power series in
the generalized Heisenberg algebra $\tilde{\mathcal{H}}_n$ defined by \eqref{1.48A}. In this case the realization of $M_{\mu\nu}$ is given by Eq. \eqref{2.53-B}
and the realization of the quantum angles $\Lambda_{\mu\nu}$ has the same form as in the orthogonal case, i.e. $\hat \Lambda_{\mu\nu}=(e^\partial)_{\mu\nu}$.
However, the powers of the matrix $\p=[\p_{\mu\nu}]$, where $\p_{\mu\nu}\in \tilde{\mathcal{H}}_n$, are computed according to
\begin{equation}\label{3.70}
(\p^0)_{\mu\nu}=\eta_{\mu\nu}, \quad (\p^m)_{\mu\nu} = \sum_{\al=1}^n \sum_{\beta=1}^n (\p^{m-1})_{\mu\al}\, \eta_{\al\beta}\, \p_{\beta\nu}, \quad m\geq 1.
\end{equation}

\subsection{Realization of the extended Poincar\'{e} algebra}

We complete our discussion by giving a brief sketch of the realization of the Poincar\'{e} algebra $\mathcal{P}=(P_\mu,M_{\mu\nu})$ and its extension by quantum angles
$\Lambda_{\mu\nu}$. Recall that the classical Poincar\'{e} algebra is defined by
\begin{align}
[P_\mu,P_\nu] &= 0,  \label{4.1}  \\
[M_{\mu\nu},P_\lambda] &= \eta_{\nu\lambda} P_\mu - \eta_{\mu\lambda} P_\nu,  \label{4.2}  \\
[M_{\mu\nu},M_{\rho\sigma}] &= \eta_{\nu\rho} M_{\mu\sigma} - \eta_{\mu\rho} M_{\nu\sigma}-\eta_{\nu\sigma} M_{\mu\rho}+\eta_{\mu\sigma} M_{\nu\rho}, \quad 1\leq \mu,\nu\leq n.  \label{4.3}
\end{align}
A realization of the algebra \eqref{4.1}--\eqref{4.3} can be constructed as follows. First, we extend the algebra $\tilde{\mathcal{H}}_n$ by adding
to relations \eqref{1.48A} $n$ pairs of generators $(p_\mu,\p_\mu)$, $1\leq \mu\leq n$, such that
\begin{equation}
[p_\mu,p_\nu] = 0, \quad [\p_\mu,\p_\nu]=0, \quad [\p_\mu,p_\nu]=\eta_{\mu\nu}
\end{equation}
with all cross--commutators being zero. Let $\tilde \KK$ be an upper--triangular matrix
\begin{equation}
\tilde \KK = \begin{bmatrix} A & 0 \\ C & D  \end{bmatrix}
\end{equation}
with blocks $A$, $C$ and $D$ defined by
\begin{align}
A_{\mu\nu} &= \p_{\mu\nu},  \\
C_{(\mu \nu) \al} &= \eta_{\al\mu} \p_\nu - \eta_{\al\nu} \p_\mu,  \\
D_{(\mu \nu) (\al\beta)} &= \frac{1}{2} \big(\eta_{\mu\al}\, \p_{\nu\beta} - \eta_{\mu\beta}\, \p_{\nu\al} + \eta_{\nu\beta}\, \p_{\mu\al}
-\eta_{\nu\al}\, \p_{\mu\beta}).
\end{align}
Then the Poincar\'{e} algebra $\mathcal{P}$ admits the realization
\begin{align}
\hat P_\mu &= \sum_{\al=1}^n \sum_{\al^\prime =1}^n p_\al \, \eta_{\al \al^\prime}\, \psi(\tilde \KK)_{\mu\al^\prime},  \label{4.4} \\
\hat M_{\mu\nu} &= \sum_{\al,\beta=1}^n \sum_{\al^\prime, \beta^\prime = 1}^n x_{\al\beta}\, \eta_{\alpha\alpha^\prime}\, \eta_{\beta\beta^\prime}\, \psi(\tilde \KK)_{(\mu \nu)(\al^\prime \beta^\prime)}
+\sum_{\al=1}^n \sum_{\al^\prime = 1}^n p_\al \eta_{\al\, \al^\prime}\, \psi(\tilde \KK)_{(\mu \nu) \al^\prime}.  \label{4.5}
\end{align}
We note that the realization \eqref{4.5} differs from the realization \eqref{2.53-B} by an extra term involving the
generators $p_\al$. The powers of the matrix $\tilde \KK$ appearing in the generating function $\psi(t)=\sum_{k=0}^\infty \frac{(-1)^k}{k!} B_k\, t^k$
are computed as follows. Zeroth power of $\tilde \KK$ is given by
\begin{equation}
\tilde \KK^0 = \begin{bmatrix} A^0 & 0 \\ 0 & D^0 \end{bmatrix}
\end{equation}
where $(A^0)_{\mu\nu} = \eta_{\mu\nu}$ and $(D^0)_{(\mu\nu)(\al\beta)} = \frac{1}{2} ( \eta_{\mu\al}\, \eta_{\nu\beta}-\eta_{\mu\beta}\, \eta_{\nu\al})$.
Since $\tilde \KK$ upper--triangular, higher powers of $\tilde \KK$ are given by
\begin{equation}
\tilde \KK^m = \begin{bmatrix} A^m & 0 \\ \sum_{k=0}^{m-1} D^k C A^{m-k-1} & D^m \end{bmatrix}, \quad m\geq 1,
\end{equation}
where the elements of each block are computed recursively according to
\begin{align}
(A^m)_{\mu\nu} &= \sum_{\al,\beta=1}^n (A^{m-1})_{\mu\al}\, \eta_{\al\beta}\, A_{\beta\nu},   \\
(D^m)_{(\mu\nu)(\al\beta)} &= \sum_{\lambda,\rho=1}^n \sum_{\lambda^\prime, \rho^\prime=1}^n (D^{m-1})_{(\mu\nu)(\lambda\rho)}\, \eta_{\lambda\lambda^\prime}\,
\eta_{\rho\rho^\prime}\, D_{(\lambda^\prime \rho^\prime)(\al\beta)}
\end{align}
and
\begin{multline}
(D^k C A^{m-k-1})_{(\mu\nu)\al} = \\ \sum_{\al^\prime, \al^{\prime\prime}=1}^n \sum_{\beta^\prime, \beta^{\prime\prime}=1}^n \sum_{\lambda^\prime, \lambda^{\prime\prime}=1}^n
(D^k)_{(\mu\nu)(\al^\prime\beta^\prime)}\, \eta_{\al^\prime\al^{\prime\prime}}\, \eta_{\beta^\prime \beta^{\prime\prime}}\,
C_{(\al^{\prime\prime} \beta^{\prime\prime}) \lambda^{\prime}}
\eta_{\lambda^\prime \lambda^{\prime\prime}}\,  (A^{m-k-1})_{\lambda^{\prime\prime} \al}.
\end{multline}

The Poincar\'{e} algebra $\mathcal{P}$ can also be extended by quantum angles $\Lambda_{\mu\nu}$ if we define (see Ref. \onlinecite{Lukierski-2})
\begin{align}
[M_{\mu\nu},\Lambda_{\rho\sigma}] &= \eta_{\nu\rho}\, \Lambda_{\mu\sigma} - \eta_{\mu\rho}\, \Lambda_{\nu\sigma},   \\
[P_\mu,\Lambda_{\rho\sigma}] &= 0,   \\
[\Lambda_{\mu\nu}, \Lambda_{\rho\sigma}] &=0.
\end{align}
In this case the realization of $\Lambda_{\mu\nu}$ is given by the exponential function $\hat \Lambda_{\mu\nu} = (e^\p)_{\mu\nu}$ where the powers of the matrix
$\p=[\p_{\mu\nu}]$ are computed recursively using Eq. \eqref{3.70}. \\

\section{Appendix}

\begin{proposition}\label{A-1}
The powers of $\KK$ are polynomials in the generators $\p_{\mu\nu}\in \mathcal{H}_n$ given by
\begin{equation}\label{App.45}
(\KK^m)_{(\mu\nu)(\al\beta)} = \frac{1}{2}\sum_{k=0}^m \binom{m}{k} \Big((\p^k)_{\mu\al} (\p^{m-k})_{\nu\beta}-
(\p^{m-k})_{\mu\beta} (\p^k)_{\nu\al}\Big), \quad m\geq 0,  \tag{A1}
\end{equation}
where $\p=[\p_{\al\beta}]$ is the $n\times n$ matrix with elements $\p_{\al\beta}$.
\end{proposition}

\proof We prove the claim by induction on $m$. Since $\KK_{(\mu\nu)(\al\beta)}$ is given by Eq. \eqref{1.30A},
one easily checks that Eq. \eqref{App.45} agrees with Eq. \eqref{1.30A} for $m=1$ where $(\p^0)_{\mu\nu}=\delta_{\mu\nu}$.
Now suppose that Eq. \eqref{App.45} holds for some $m>1$. Then it follows from the induction hypothesis that
\begin{align}
&(\KK^{m+1})_{(\mu\nu)(\al\beta)} = \sum_{\theta,\sigma=1}^n (\KK^m)_{(\mu\nu)(\theta\sigma)} \KK_{(\theta\sigma)(\al\beta)} \notag \\
&=\frac{1}{4}\sum_{\theta,\sigma=1}^m \sum_{k=0}^m \binom{m}{k} \Big((\p^k)_{\mu\theta}\, (\p^{m-k})_{\nu\sigma} - (\p^{m-k})_{\mu\sigma}\,
(\p^k)_{\nu\theta}\Big) (\delta_{\theta\al} \p_{\sigma\beta} - \delta_{\theta\beta} \p_{\sigma\al} + \delta_{\sigma\beta} \p_{\theta\al} -
\delta_{\sigma\al} \p_{\theta\beta}) \notag \\
&=\frac{1}{4} \sum_{k=0}^m \binom{m}{k} \Big((\p^k)_{\mu\al}\, (\p^{m+1-k})_{\nu\beta} - (\p^{k+1})_{\mu\beta}\, (\p^{m-k})_{\nu\al}+(\p^{k+1})_{\mu\al}\,
(\p^{m-k})_{\nu\beta} \notag \\
&-(\p^k)_{\mu\beta}\, (\p^{m+1-k})_{\nu\al} + (\p^{m-k})_{\mu\al}\, (\p^{k+1})_{\nu\beta}-(\p^{m+1-k})_{\mu\beta}\, (\p^k)_{\nu\al} \notag \\
&+ (\p^{m+1-k})_{\mu\al}\, (\p^k)_{\nu\beta} - (\p^{m-k})_{\mu\beta}\, (\p^{k+1})_{\nu\al}\Big).    \tag{A2}
\end{align}
Using
\begin{equation}
\sum_{k=0}^m \binom{m}{k} x^k y^{m-k} = \sum_{k=0}^m \binom{m}{k} x^{m-k} y^k   \tag{A3}
\end{equation}
and collecting terms with the same pairs of indices we find
\begin{align}
(\KK^{m+1})_{(\mu\nu)(\al\beta)} &= \frac{1}{2} \sum_{k=0}^m \binom{m}{k} \Big((\p^k)_{\mu\al}\, (\p^{m+1-k})_{\nu\beta}-
(\p^{m+1-k})_{\mu\beta}\, (\p^k)_{\nu\al} \notag \\
&+(\p^{k+1})_{\mu\al}\, (\p^{m-k})_{\nu\beta}-(\p^{m-k})_{\mu\beta}\, (\p^{k+1})_{\nu\al}\Big) \notag \\
&=\frac{1}{2} \sum_{k=0}^m \binom{m}{k} \Big( (\p^k)_{\mu\al}\, (\p^{m+1-k})_{\nu\beta}-(\p^{m+1-k})_{\mu\beta}\, (\p^k)_{\nu\al}\Big) \notag \\
&+\frac{1}{2}\sum_{k=1}^{m+1} \binom{m}{k-1} \Big((\p^k)_{\mu\al}\, (\p^{m+1-k})_{\nu\beta} - (\p^{m+1-k})_{\mu\beta}\, (\p^k)_{\nu\al}\Big) \notag \\
&=\frac{1}{2} \sum_{k=0}^{m+1} \binom{m+1}{k} \Big((\p^k)_{\mu\al}\, (\p^{m+1-k})_{\nu\beta} - (\p^{m+1-k})_{\mu\beta}\, (\p^k)_{\nu\al}\Big)   \tag{A4}
\end{align}
where the last equality follows from the binomial identity
\begin{equation}
\binom{m}{k}+\binom{m}{k-1} = \binom{m+1}{k}.   \tag{A5}
\end{equation}
This proves the claim. \qed

\begin{proposition}\label{A-3}
Let $\p=[\p_{\mu\nu}]$ be the matrix of generators $\p_{\mu\nu}\in \mathcal{H}_n$ satisfying the commutation relations \eqref{Hn}. Then
\begin{equation}\label{App. 102}
[x_{\al\beta},(\p^m)_{\rho\sigma}] = 2\sum_{k=1}^m \binom{m}{k} \sum_{\mu=1}^n (-\KK)^{k-1}_{(\al\beta)(\mu\rho)} \big(\p^{m-k}\big)_{\mu\sigma}, \quad m\geq 1.  \tag{A6}
\end{equation}
\end{proposition}

\proof We rewrite the right--hand side of Eq. \eqref{App. 102} as a polynomial in $\p_{\mu\nu}$ and prove the claim by induction on $m$. Since the powers of $\KK$ are given by
Eq. \eqref{App.45}, by straightforward but lengthly computation we find
\begin{align}
&2\sum_{k=1}^m \binom{m}{k} \sum_{\mu=1}^n (-\KK)^{k-1}_{(\al\beta)(\mu\rho)}\, (\p^{m-k})_{\mu\sigma} \notag \\
&=\sum_{k=1}^m \sum_{l=0}^{k-1}
(-1)^{k-1} \binom{m}{k} \binom{k-1}{l} \Big[\big(\p^{m-(k-l)}\big)_{\al\sigma}\, \big(\p^{k-l-1}\big)_{\beta\rho}-\big(\p^{k-l-1}\big)_{\al\rho}\,
\big(\p^{m-(k-l)}\big)_{\beta\sigma}\Big] \notag \\
&=\sum_{p=1}^m \sum_{\substack{1\leq k\leq m \\ 0 \leq l \leq m-1 \\k-l=p}} (-1)^{k-1} \binom{m}{k} \binom{k-1}{l} \Big[(\p^{m-p})_{\al\sigma}\, (\p^{p-1})_{\beta\rho} -
(\p^{p-1})_{\al\rho}\, (\p^{m-p})_{\beta\sigma}\Big]  \label{App. 103}  \tag{A7}
\end{align}
where the inner sum runs over all indices $1\leq k\leq m$ and $0\leq l \leq m-1$ such that $k-l=p$. In view of the  identity
\begin{equation}
\sum_{k-l=p} (-1)^{k-1} \binom{m}{k} \binom{k-1}{l} = (-1)^{p-1},  \tag{A8}
\end{equation}
Eq. \eqref{App. 103} simplifies to
\begin{equation}
2\sum_{k=1}^m \binom{m}{k} \sum_{\mu=1}^n (-\KK)^{k-1}_{(\al\beta)(\mu\rho)}\, (\p^{m-k})_{\mu\sigma} =
\sum_{p=1}^m (-1)^{p-1}\Big[ (\p^{m-p})_{\al\sigma}\, (\p^{p-1})_{\beta\rho}-(\p^{p-1})_{\al\rho}\, (\p^{m-p})_{\beta\sigma}\Big]  \tag{A9}
\end{equation}
Therefore, we need to prove that
\begin{equation}\label{App. 106}
[x_{\al\beta},(\p^m)_{\rho\sigma}] = \sum_{p=1}^m (-1)^{p-1}\Big[ (\p^{m-p})_{\al\sigma}\, (\p^{p-1})_{\beta\rho}-(\p^{p-1})_{\al\rho}\, (\p^{m-p})_{\beta\sigma}\Big] \tag{A10}
\end{equation}
for all $m\geq 1$.
Since $[x_{\al\beta},\p_{\rho\sigma}]=\delta_{\rho\beta}\, \delta_{\sigma\al}-\delta_{\rho\al}\, \delta_{\sigma\beta}$, one easily verifies that Eq. \eqref{App. 106} holds
for $m=1$. Now suppose that Eq. \eqref{App. 106} holds for some $m>1$. Then, using Eq. \eqref{App. 106} and noting that $(\p^m)_{\beta\al} = (-1)^m (\p^m)_{\al\beta}$ since
$\p$ is an anti--symmetric matrix we have
\begin{align}
[x_{\al\beta},(\p^{m+1})_{\rho\sigma}] &= \sum_{\mu=1}^n [x_{\al\beta},(\p^m)_{\rho\mu}]\, \p_{\mu\sigma} + \sum_{\mu=1}^n (\p^m)_{\rho\mu}\, [x_{\al\beta}, \p_{\mu\sigma}] \notag \\
&=\sum_{p=1}^m (-1)^{p-1} \big[(\p^{m+1-p})_{\al\sigma} \, (\p^{-1})_{\beta\rho}-(\p^{p-1})_{\al\rho}\, (\p^{m+1-p})_{\beta\sigma}\big] \notag \\
&+(-1)^m \big[(\p^0)_{\al\sigma}\, (\p^m)_{\beta\rho}-(\p^m)_{\al\rho}\, (\p^0)_{\beta\sigma}\big] \notag \\
&=\sum_{p=1}^{m+1} (-1)^{p-1}\big[(\p^{m+1-p})_{\al\sigma}\, (\p^{p-1})_{\beta\rho}-(\p^{p-1})_{\al\rho}\, (\p^{m+1-p})_{\beta\sigma}\big].  \tag{A11}
\end{align}
By induction, Eq. \eqref{App. 106} holds for all $m\geq 1$. \qed

\subsection*{Data Availability Statement}

\noindent Data sharing is not applicable to this article as no new data were created or analyzed in this study.

%%%%%%%%%%%%%%%%%%%%%%%%%%%%%%%%%%%%%%%%%%%%%%%%%%%%%%%%%%%%%%%%%%%%%%%%%%%%%%%%%%%%%%%%%%%%%%%%%%%%%%%%%%%%%%%%%%%%%%%%%%%%%%%%%%%%%%%%%%%%%%%%%%%%%%%%%%%%%%%%%%%%%%%%%%%%%%%%%%%%%%%%%%%

\end{document}